\documentclass[9pt,twocolumn,twoside]{osajnl}
\journal{ol} 

\setboolean{shortarticle}{true}

\ifthenelse{\boolean{shortarticle}}{\colorlet{color2}{color2}}{\colorlet{color2}{color2}} 

\title{Stationary and oscillatory bound states of dissipative solitons
created by the third-order dispersion}

\title{Stationary and oscillatory bound states of dissipative solitons
created by the third-order dispersion}

\author[1,*]{Hidetsugu Sakaguchi}
\author[2,3]{Dmitry V. Skryabin}
\author[3,4] {Boris A. Malomed}

\affil[1]{Department of Applied Science for Electronics and Materials, Interdisciplinary Graduate School of Engineering Sciences, Kyushu University, Kasuga, Fukuoka 816-8580, Japan}
\affil[2]{Department of Physics, University of Bath, Bath, BA2 7AY, UK}
\affil[3]{ITMO University, St. Petersburg, 197101, Russia}
\affil[4]{Department of Physical Electronics, School of Electrical Engineering, Faculty of Engineering, Tel Aviv University, P.O.B. 39040, Ramat Aviv, Tel Aviv, Israel}

\affil[*]{Corresponding author: sakaguchi@asem.kyushu-u.ac.jp}

\dates{Compiled \today}

\ociscodes{(190.5530) Nonlinear optics, pulse propagation and temporal solitons; (060.4370) Nonlinear optics, fibers; (140.3510) Lasers, fiber; (190.3100) Nonlinear optics, instabilities and chaos}

\begin{abstract}
We consider the model of fiber-laser cavities near the zero-dispersion point, based on the complex Ginzburg-Landau equation with the cubic-quintic nonlinearity and including third-order dispersion (TOD)
term. It is well known that this model supports stable dissipative solitons.
We demonstrate that the same model gives rise to several families of robust bound states of the solitons, which exists only in the presence of the TOD. There are both stationary and dynamical bound states, with
oscillating separation between the bound solitons. Stationary states are multistable, corresponding to
different values of the separation. With the
increase of the TOD coefficient, the bound state with the smallest separation gives rise the oscillatory
state through
the Hopf bifurcation. Further growth of TOD leads to a bifurcation transforming the oscillatory limit cycle into a strange attractor, which represents the chaotically oscillating dynamical bound state. Families of multistable three- and four-soliton complexes are found too, the ones with the smallest separation between the solitons
again ending by the transition to oscillatory states through the Hopf bifurcation.
\end{abstract}
\setboolean{displaycopyright}{true}

\begin{document}

\maketitle

Experimental and theoretical studies of temporal solitons in fiber lasers is
a vast area of fundamental and applied research 
\cite{KA}-\cite{AA},
which suggests generic paradigms of stable dynamics of dissipative solitons
to many other fields, in optics and beyond \cite{Rosanov,Mihalache}. In
addition to isolated solitons, experiments and theoretical (analytical and
numerical) investigations of models based on complex Ginzburg-Landau
equations (CGLEs) reveal the existence of stable bound complexes of
dissipative fiber solitons, starting from the prediction \cite{1991} and
first experimental observation \cite{Tang2}. The latter subject has drawn a
great deal of interest in the course of the last two decades \cite%
{Tang,Mitschke,Tang2,Angers}.

The fundamental model of lossless nonlinear fibers based on the cubic
nonlinear Schr\"{o}dinger equation (NLSE)\ does not admit solutions in the
form of bound states. The simplest possibility to produce them is to include
the third-order dispersion (TOD), represented by the linear term with the
third derivative. It is well known that this modification of the NLSE gives
rise to bound states, which are, however, structurally unstable \cite%
{Akylas,Jianke
}. The TOD term is a relevant one in fiber-optic
settings when the pump wavelength is close to the zero-GVD (group-velocity
dispersion) point. While dealing with fiber lasers, it is also imperative to
include gain and loss, the balance between which makes it possible to create
stable dissipative solitons. A necessary condition for the stability of
bright solitons is the stability of their background, which, in turn, makes
one to seek soliton lasing regimes below the continuous-wave lasing
threshold. Mathematically it implies that the CGLE should include the cubic
gain and overall quintic loss. The corresponding cubic-quintic (CQ)
nonlinearity, which was first phenomenologically proposed by Petviashvili
and Sergeev \cite{Petvia}, can be derived as an approximate fiber-laser
model, which is commonly and successfully used \cite{sat-abs,Soto,Leblond}.
The existence of stable dark \cite{Hidetsugu} and bright \cite{appendix}-%
\cite{Chate} dissipative solitons in the CGLE, as well as of their bound
states \cite{Seva}, has been firmly established in theory and experimentally.

The form and stability of isolated bright dissipative solitons in the
CQ-CGLE with the TOD term was investigated too, but to a lesser degree \cite%
{CQTOD1,CQTOD2,CQTOD3}. A possibility of the existence of the soliton bound
states in this model is a natural extension of the analysis, with obvious
perspectives for experimental realization and applications. To address this
possibility, we adopt the scaled CQ-CGLE for complex amplitude $u\left(
z,\tau \right) $ of the electromagnetic field, where $z$ and $\tau $ are the
propagation distance and reduced time \cite{KA}:%
\begin{equation}
i\frac{\partial u}{\partial z}=\left(i\delta -\beta _{2}\right) \frac{%
\partial ^{2}u}{\partial \tau ^{2}}-|u|^{2}u - i\left( \varepsilon -\alpha
_{1}|u|^{2}+\alpha _{2}|u|^{4}\right) u+i\beta _{3}\frac{\partial ^{3}u}{%
\partial \tau ^{3}}.  \label{CGLE}
\end{equation}%
Here $\beta _{2}$ and $\beta _{3}$ are, respectively, the usual second-order
GVD and TOD coefficients ($\beta _{2}>0$ corresponds to the anomalous GVD), $%
\delta \geq 0$ is the spectral-filtering parameter (dispersive losses), the
effective Kerr coefficient is normalized to be $1$, and positive constants $%
\varepsilon $, $\alpha _{1}$, and $\alpha _{2}$ account for the linear loss,
cubic gain, and quintic loss, respectively.

First, to forecast the existence of bound states of dissipative solitons, it
is relevant to analyze the structure of their tails, which overlap with
cores of neighboring solitons, giving rise to the effective interaction
potential, $U(T)$, where $T$ is the temporal separation between the cores.
Local minima of the potential, if any, predict values of $T$ for stationary
bound states \cite{1991}. The solution for the decaying tail of a
dissipative soliton with a propagation constant, $k$, is looked for as
\begin{equation}
u\left( z,\tau \right) =u_{0}\exp \left( ikz-\left( \chi +i\omega \right)
|\tau |-i\Omega \tau \right) ,  \label{tail}
\end{equation}%
where $1/\chi >0$ is the temporal width of the soliton, $\omega $ determines
oscillations of the tails on both sides of the soliton, and $\Omega $ is an
overall frequency shift. Assuming $\delta $, $\varepsilon $, and $\beta _{3}$
to be small perturbations, one can solve the resulting cubic equations for
real constants $\chi $, $\omega $, and $\Omega $:%
\begin{equation}
\chi \approx \sqrt{\frac{k}{\beta _{2}}},\omega \approx \frac{\delta
k-\varepsilon \beta _{2}}{2\beta _{2}^{3/2}\sqrt{k}},\Omega \approx -\frac{%
\beta _{3}k}{2\beta _{2}^{2}},  \label{k}
\end{equation}%
where $\beta _{2}k>0$ is assumed. Note that Eq. (\ref{k}) makes it possible
to predict, in the present approximation, the group velocity induced by the
TOD term, as per Eq. (\ref{k}):
\begin{equation}
v\equiv \left( d\tau /dz\right) _{\mathrm{gr}}=d\Omega /dk\approx -\left(
2\beta _{2}^{2}\right) ^{-1}\beta _{3}.  \label{gr}
\end{equation}

An essential peculiarity of the tail solution (\ref{tail}) is the
combination of $\tau $ and $|\tau |$ in the phase of Eq. (\ref{tail}). Then,
following a standard method \cite{1991} for constructing the effective
interaction potential, $U(T,\phi )$, for the pair of solitons with phase
shift $\phi $ and temporal separation $T$, one obtains
\begin{equation}
U\left( T,\phi \right) =U_{0}\exp \left( i\phi -\chi T\right) \cos \left(
\omega T\right) \cos \left( \Omega T\right) .  \label{U}
\end{equation}%
Equation (\ref{U}) gives rise to the two sets of the soliton separation
distances,
\begin{equation}
T_{m}^{(\omega )}=\left( \pi /2\omega \right) \left( 1+2m\right)
;~T_{n}^{(\Omega )}=\left( \pi /2\Omega \right) \left( 1+2n\right) ,
\label{T}
\end{equation}%
with integer $m$ and $n$, corresponding to potential minima and hence to
stationary bound states. Note that $\Omega \sim \beta _{3}$ in Eq. (\ref{k})
implies that TOD introduces a new family of bound states that are absent in
the well studied no-TOD case.

Numerical results are produced here for fixed parameters
\begin{equation}
\beta _{2}=0.1,\delta =0.01,\varepsilon =0.01,\alpha _{1}=0.06,\alpha
_{2}=0.006,  \label{fixed}
\end{equation}%
while varying the TOD coefficient, $\beta _{3}$, as broader numerical data
demonstrate that this parameter set adequately represents the generic
situation, the deviation from Eq. (\ref{fixed}) producing inconspicuous
changes in the results (see details below). In terms of physical parameters,
these values correspond, roughly, to fiber solitons with temporal width $%
\sim 1$ ps.

\begin{figure}[tbh]
\vspace{-0.3cm} \centering
{\includegraphics[width=\linewidth]{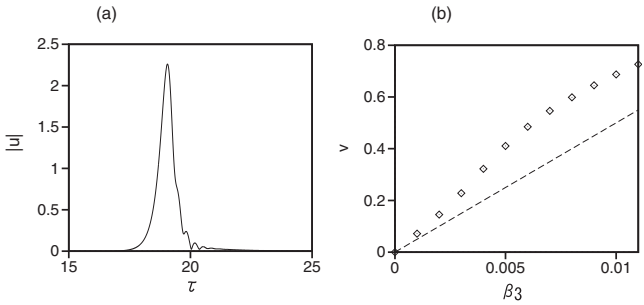}}
\caption{(a) A typical profile of a stable dissipative solitons, produced by
the numerical solution of Eq. (\protect\ref{CGLE}), for parameters given by
Eq. (\protect\ref{fixed}) and $\protect\beta _{3}=0.01$. (b) The soliton's
velocity vs. $\protect\beta _{3}$, the dashed line showing the prediction
given by Eq.\ (\protect\ref{gr}). }
\label{fig1}
\end{figure}

\begin{figure}[h]
\centering{\includegraphics[width=\linewidth]{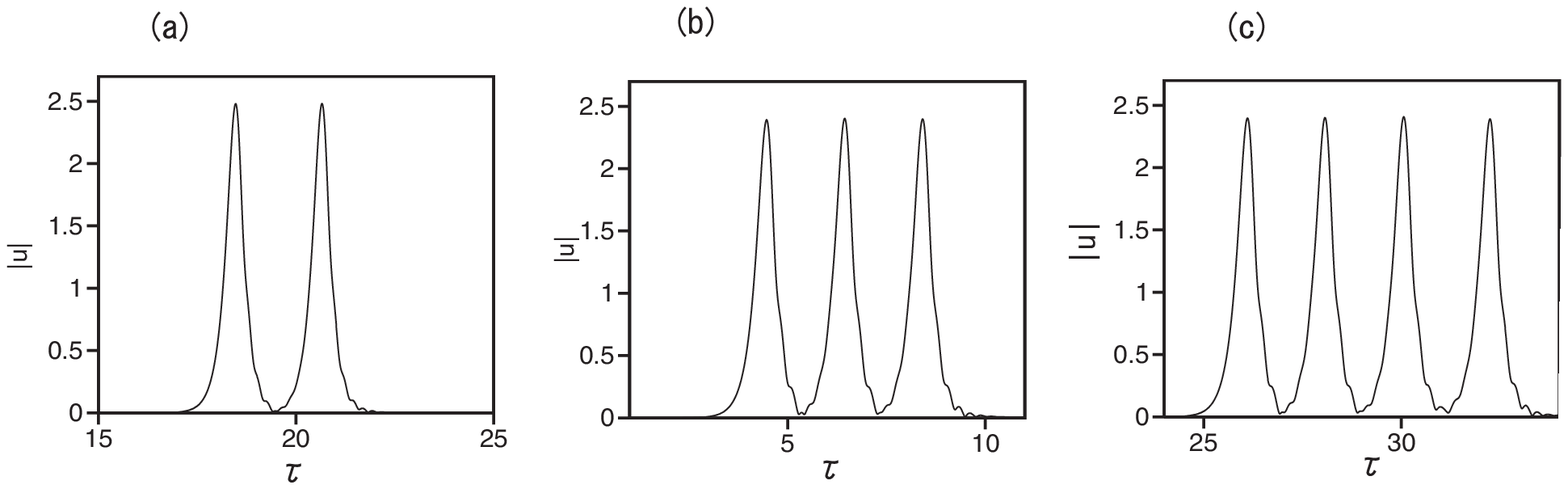}}
\caption{Stable static bound states: (a) a two-soliton one at $\protect\beta%
_3=0.005$; (b) and (c) three- and four-soliton complexes found at $\protect%
\beta_3=0.0065$. Here and below, other parameters are fixed as per Eq. (%
\protect\ref{fixed}).}
\label{fig2}
\end{figure}

\begin{figure}[tbh]
\centering
{\includegraphics[width=\linewidth]{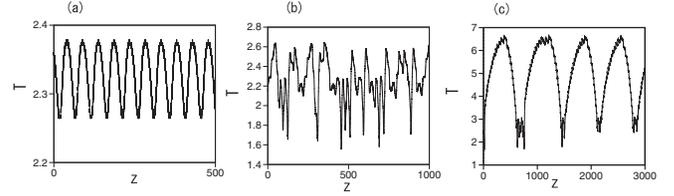}}
\caption{The separation between solitons in stable two-soliton oscillatory
bound states vs. $z$, at (a) $\protect\beta_3=0.009$, (b) $\protect\beta%
_3=0.011$, and (c) $\protect\beta_3=0.02$ The panels represent typical
examples of regular, chaotic, and large-amplitude quasiperiodic oscillations.
}
\label{fig3}
\end{figure}

The numerical solution of Eq. (\ref{CGLE}) reveals, along with single
solitons (see Fig. \ref{fig1}), their bound states of two types, \textit{viz}%
., \emph{static} ones with $z$-independent separation $T$ (Fig. \ref{fig2}),
and \emph{dynamical} bound states, with $T$ oscillating in $z$, as shown in
Fig. \ref{fig3}.

In Fig. \ref{fig1}(a), the soliton's profile $|u(\tau )|$, traveling at
constant velocity $d\tau /dz=0.687$, is shown for $\beta _{3}=0.01$. The
front side of the pulse features an oscillating tail, which is reasonably
well approximated by Eq. (\ref{U}), while the trailing tail decays
monotonously. Figure \ref{fig1}(b) shows the velocity $v$ as a function of $%
\beta _{3}$, and its approximation provided by Eq. (\ref{gr}), which does
not take into account nonlinear and dissipative terms in Eq. (\ref{CGLE}),
but nevertheless produces a reasonable approximation. For complexes of bound
solitons, dependence $v(\beta _{3})$ is very similar to that displayed in
Figure \ref{fig1}(b). Solitons bound in static states always have equal
amplitudes, while in the oscillatory state the relative difference between
instantaneous amplitudes may be \ $\sim 10\%$.

Figure \ref{fig3} shows that the two-soliton dynamical bound states exhibit
both periodic and chaotic oscillations. Generally, the increase of the TOD
parameter, $\beta _{3}$, leads to the transition from static bound states to
regularly oscillating ones, and further to ones with random internal
oscillations. Further, Fig. \ref{fig3}(c) demonstrates quasiperiodic
oscillations observed at still larger $\beta _{3}$, with a superposition of
long-period large-amplitude oscillations and fast small-amplitude vibrations
of the two-soliton states.

\begin{figure}[tbh]
\centering
{\includegraphics[width=\linewidth]{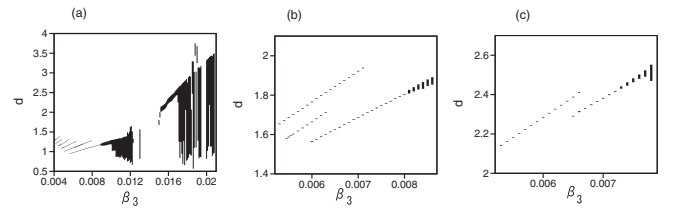}}
\caption{(a) The separation between two bound solitons vs. the TOD
coefficient, $\protect\beta _{3}$. Vertical segments represent oscillations
of the separation in dynamical bound states. Multiple tilted dotted lines
correspond to coexisting stationary bound stated with different separations.
(b,c) The same for the temporal distance between adjacent solitons in three
and four-soliton complexes, respectively.}
\label{fig4}
\end{figure}

Results produced by the systematic simulations of the two-soliton bound
states are summarized in Fig. \ref{fig4}(a), which shows the separation
between solitons in stable bound states as a function of the TOD
coefficient, $\beta _{3}$. In this plot, a single dot marks a separation
between the core in a static bound state of two solitons. Each oscillatory
state, periodic, chaotic, or quasiperiodic one, is represented by a vertical
segment which covers an interval of values of the separation covered by the
oscillations. In broad white gaps, no stable static or dynamical bound
states were produced by the simulations. It is worthy to note that the
presence of the TOD is indeed necessary for the existence of the bound
states represented in Fig. \ref{fig4}, as their families emerge at finite
values of $\beta _{3}$.

A noteworthy finding observed in Fig. \ref{fig4}(a) is the multistability of
the static bound states with different values of the temporal separation
between the paired solitons. In accordance with the equidistant spectra of
the separation, predicted by Eq. (\ref{k}), differences between values of
the separation in the coexisting states are approximately equal. In the case
shown in Fig. \ref{fig4}, six distinct families of the stationary states are
found, demonstrating bistability and tristability: two or three different
stable bound states may coexist at given values of parameters. On the other
hand, stationary bound states never coexist with dynamical ones.

Another essential feature seen in Fig. \ref{fig4}(a) is that, while the
static states corresponding to higher values of the separation terminate
with the increase of $\beta _{3}$, the branch corresponding to the minimum
separation undergoes a Hopf bifurcation, which replaces the stable fixed
point by an emerging limit cycle \cite{bif}, at $\beta _{3}\approx 0.009$.
The subsequent transition from regular to chaotic oscillations may be
understood as a bifurcation which transforms the limit cycle into a strange
attractor \cite{bif}.

The existence and stability of the bound states reported here most
essentially depends on the TOD coefficient, $\beta _{3}$. The dependence on
other parameters was explored too, showing that the bound states are weakly
sensitive to their variations. For instance, the variation of the
linear-loss parameter $\varepsilon $ in Eq. (\ref{CGLE}) by a factor of $%
\simeq 10$ changes the value of $\beta _{3}$ at the boundary between the
static and oscillatory states only by $\simeq 25\%$.

The presence of stable paired states of two dissipative solitons suggests
looking for multi-soliton complexes, which were also observed experimentally
in fiber lasers \cite{tri,multi}. Indeed, in the same parameter region where
multistable two-soliton states are found, three- and four-soliton complexes
are present too, as shown in Fig. \ref{fig2}(b,c), with equal separations
between adjacent solitons. Further, Figs. \ref{fig4} demonstrates that the
multi-soliton states are multistable too, and the one with the smallest
separation between the bound solitons, similarly its two-soliton
counterpart, ends by a Hopf bifurcation, which leads to a robust complex
with periodic oscillations of the separation between the bound solitons.

In conclusion, by means of an analytical approximation and systematic
numerical calculations, we have demonstrated that the addition of the TOD
term to the standard model of fiber-laser cavities, based on the CGLE with
the CQ nonlinearity, which is an essential ingredient of the model near the
zero-dispersion point, creates families of stable two-, three-, and
four-soliton bound states, which do not exist in the absence of TOD. The
bound states may be stationary and oscillatory, including robust bound pairs
of two dissipative solitons with the chaotically or quasiperiodically
oscillating separation between them. A noteworthy finding is the
multistability of stationary two- and multi-soliton bound states, which
correspond to different separations between the solitons.

Funding: Royal Society (IE 160465); Israel Science Foundation (1286/17). EU
H2020 (691011-Soliring); ITMO University Visiting Professorship via the
Government of Russia Grant 074-U01; Russian Foundation for Basic Research
(17-02-00081).


\begin{thebibliography}{99}

\bibitem{KA} Y. S. Kivshar and G. P. Agrawal, \textit{Optical Solitons: From
Fibers to Photonic Crystals} (Academic Press: San Diego, 2003)

\bibitem{Wise} F. W. Wise, A. Chong and W. H. Renninger, High-energy
femtosecond fiber lasers based on pulse propagation at normal dispersion,
Laser Phot. Rev. \textbf{2}, 58-73 (2008).

\bibitem{AA} A. Ankiewicz and N. Akhmediev (Eds.), \textit{Dissipative
Solitons: From Optics to Biology and Medicine} (Springer: Heidelberg, 2008).

\bibitem{Rosanov} N. N. Rosanov, \textit{Spatial Hysteresis and Optical
Patterns} (Springer, 2002).

\bibitem{Mihalache} D. Mihalache, Localized structures in nonlinear optical
media: A selection of recent studies, Rom. Rep. Phys. \textbf{67}, 1383-1400
(2015).

\bibitem{1991} B. A. Malomed, Bound solitons in the nonlinear Schr\"{o}%
dinger - Ginzburg-Landau equation. Phys. Rev. A \textbf{44}, 6954-6956
(1991).

\bibitem{Tang} D. Y. Tang, W. S. Man, H. Y. Tam, and P. D. Drummond,
Observation of bound states of solitons in a passively mode-locked fiber
laser, Phys. Rev. A \textbf{64}, 033814 (2001).

\bibitem{eight} N. H. Seong and D. Y. Kim, Experimental observation of
stable bound solitons in a figure-eight fiber laser, Opt. Lett. \textbf{27},
1321-1323 (2002).

\bibitem{Mitschke} M. Stratmann, T. Pagel, and F. Mitschke, Experimental
observation of temporal soliton molecules, Phys. Rev. Lett. \textbf{95},
143902 (2005).

\bibitem{Tang2} D. Y. Tang, B. Zhao, L. M., Zhao, and H. Y. Tam, Soliton
interaction in a fiber ring laser, Phys. Rev. E \textbf{72}, 016616 (2005).

\bibitem{Angers} F. Amrani, A. Haboucha, M. Salhi, H. Leblond, A. Komarov,
and F. Sanchez, Dissipative solitons compounds in a fiber laser. Analogy
with the states of the matter, Appl. Phys. B: Lasers Opt. \textbf{99},
107-114 (2010).

\bibitem{Akylas} D. C. Calvo and T. R. Akylas, Stability of bound states
near the zero-dispersion wavelength in optical fibers, Phys. Rev. E \textbf{%
56}, 4757-4664 (1997).

\bibitem{Jianke} J. Yang and T. R. Akylas, Continuous families of embedded
solitons in the third-order nonlinear Schr\"{o}dinger equation, Stud. Appl.
Math. \textbf{111}, 359-375 (2003).


\bibitem{Petvia} V. I. Petviashvili and A. M. Sergeev, Spiral solitons in
active media with excitation thresholds, Dokl. Akad. Nauk SSSR \textbf{276},
1380-1384 (1984).

\bibitem{sat-abs} W. J. Firth and A. J. Scroggie, Optical bullet holes:
Robust controllable localized states of a nonlinear cavity, Phys. Rev. Lett.
\textbf{76}, 1623-1626 (1996).

\bibitem{Soto} J. M. Soto-Crespo, N. Akhmediev, and A. Ankiewicz, Pulsating,
creeping, and erupting solitons in dissipative systems, Phys. Rev. Lett.
\textbf{85}, 2937-2940 (2000).

\bibitem{Leblond} A. Komarov, H. Leblond, and F. Sanchez, Quintic complex
Ginzburg-Landau model for ring fiber lasers, Phys. Rev. E \textbf{72},
025604 (2005).

\bibitem{appendix} B. A. Malomed, Evolution of nonsoliton and
\textquotedblleft quasiclassical\textquotedblright\ wavetrains in nonlinear
Schr\"{o}dinger and Korteweg - de Vries equations with dissipative
perturbations. Physica D \textbf{29}, 155-172 (1987).

\bibitem{Fauve} S. Fauve and O. Thual, Solitary waves generated by
subcritical instabilities in dissipative systems, Phys. Rev. Lett. \textbf{64%
}, 282-284 (1990).

\bibitem{Alik} B. A. Malomed and A. A. Nepomnyashchy, Kinks and solitons in
the generalized Ginzburg-Landau equation, Phys. Rev. A \textbf{42},
6009-6014 (1990).

\bibitem{Pomeau} V. Hakim, P. Jakobsen, and Y. Pomeau, Fronts vs. solitary
waves in nonequilibrium systems, Europhys. Lett. \textbf{11}, 19-24 (1990).

\bibitem{Hoh} W. van Saarloos and P. C. Hohenberg, Pulses and fronts in the
complex Ginzburg-Landau equation near a subcritical bifurcation. Phys. Rev.
Lett. \textbf{64}, 749-752 (1990).

\bibitem{Chate} P. Marcq, H. Chat\'{e}, R. Conte, Exact solutions of the
one-dimensional quintic complex Ginzburg-Landau equation. Physica D \textbf{%
73}, 305-317 (1994).

\bibitem{Hidetsugu} H. Sakaguchi, Hole solutions in the complex
Ginzburg-Landau equation near a subcritical bifurcation, Prog. Theor. Phys.
\textbf{86}, 7-12 (1991).

\bibitem{Seva} V. V. Afanasjev, B. A. Malomed, and P. L. Chu, Stability of
bound states of pulses in the Ginzburg-Landau equations, Phys. Rev. E
\textbf{56}, 6020-6025 (1997).

\bibitem{CQTOD1} L. Song, L. Li, Z. Li, and G. Zhou, Effect of third-order
dispersion on pulsating, erupting and creeping solitons, Opt. Commun.
\textbf{249}, 301-209 (2005).

\bibitem{CQTOD2} I. M. Uzunov, T. N. Arabadzhiev, and Z. D. Georgiev,
Influence of higher-order effects on pulsating solutions, stationary
solutions and moving fronts in the presence of linear and nonlinear
gain/loss and spectral filtering, Opt. Fiber Techn. \textbf{24}, 15-23
(2015).

\bibitem{CQTOD3} S. C. Latas, M. F. S. Ferreira, and M. Fac\~{a}o,
Ultrashort high-amplitude dissipative solitons in the presence of
higher-order effects, J. Opt. Soc. Am B \textbf{34}, 1032-1040 (2017).

\bibitem{bif} E. Ott, Chaos in Dynamical Systems (Cambridge University
Press: Cambridge, 2002).

\bibitem{tri} B. Ortac, A. Hideur, T. Chartier, M. Brunel, P. Grelu, H.
Leblond, and E. Sanchez, Generation of bound states of three ultrashort
pulses with a passively mode-locked high-power Yb-doped double-clad fiber
laser, IEEE\ Phot. Tech. Lett. \textbf{16}, 1274-1276 (2004).

\bibitem{multi} F. Amrani, M. Salhi, P. Grelu, H. Leblond, and F. Sanchez,
Universal soliton pattern formations in passively mode-locked fiber lasers,
Opt. Lett. \textbf{36}, 1545-1547 (2011).
\end{thebibliography}



\end{document}